\documentstyle[aaspp4,epsf]{article}
\def\la{\hbox{\raise.5ex\hbox{$<$}
    \kern-1.1em\lower.5ex\hbox{$\sim$}}}
\def\ga{\hbox{\raise.5ex\hbox{$>$}
    \kern-1.1em\lower.5ex\hbox{$\sim$}}}

\lefthead{Salaris, Degl'Innocenti, Weiss:}
\righthead{Globular cluster ages}

\begin{document}

\title{The age of the oldest globular clusters}
\author{M.~Salaris, S.~Degl'Innocenti and A.~Weiss}
\affil{Max-Planck-Institut f\"ur Astrophysik, Postfach 1523, 85740
Garching, Germany}

\begin{abstract}
The age of three of the oldest clusters -- M15, M68, M92 -- has been
redetermined. We use the latest EOS and opacity data available for
calculating both isochrones and zero age horizontal branches and employ the
brightness difference between turn-off and horizontal branch to
determine the cluster age. Our best ages for all three clusters are about
13 Gyr, and even smaller ages are possible.
Our results help to reconcile cluster ages with recent
results on the age of the universe determined from the Hubble
constant.
\end{abstract}


\section{Introduction}

If the age of the universe is obtained from the cosmological
expansion  by determining the Hubble constant $H_0$, it turns
out to be less than 15 Gyr for $H_0\, \ga\, 50 {\rm km/(s\, Mpc)}$ for
values of the density parameter $\Omega_0\, \ga\, 0.4$. The most recent
investigations into $H_0$ (for a review, see
van den Bergh 1994) yield values between 60 and 90, which
limit the age of the universe to less than 13 Gyr for any reasonable
$\Omega_0$, even if a cosmological constant is allowed. This number
even reduces to 10 Gyr, if $H_0 \approx 75$ (from Cepheids or
SNIa). Clearly, even if one stretches the limits, the universe appears
to be younger than $\approx 15$ Gyr.

On the other hand, the classical method of determining the age of the
oldest known stellar objects, the isochrone fitting to metal-poor
globular clusters, consistently yields ages above $\approx 14$ Gyr for
the bulk of metal-poor halo clusters, with the majority of the
clusters being around $16$ Gyr and the oldest ones up to $18$ Gyr old.
An additional Gyr has to be added to this for the time between the
genesis of the universe and the creation of the first clusters.
This ``Conflict over the age of the Universe'' has been brought to the
point by Bolte \& Hogan (1995) by the example of M92, probably the
best observed very old cluster, whose age they give as $15.8
\pm 2.1$ Gyr (this range was used by Kennicutt, Freedman \&
Mould~(1995) to illustrate the relation -- and conflict -- with
$H_0$).

It therefore is necessary to reconsider age determinations of globular
clusters and to investigate the errors more carefully. In this {\sl
Letter} we report about new age determinations of three of the oldest
clusters, including M92, using standard approaches, but latest
equation of state and opacities for the stellar models.
Cluster ages were determined by using the difference in visual
magnitude ($\Delta(V)$) between main-sequence turn-off and 
horizontal branch (HB).

Our work is similar to Chaboyer \& Kim (1995; CK95), who used the same
method and already found  a reduction of cluster ages of
about 7\% (or 1.2 Gyr for the oldest clusters). However, they had
determined $\Delta(V)$ from the difference between the theoretical
turn-off luminosity $V_{TO}$ and an empirical horizontal branch
brightness $V_{HB}$ (based on RR~Lyrae luminosities). In contrast, we
calculate theoretical ZAHBs as well and fit both isochrones and ZAHB to the
observations. This way, the influence of the new input physics is
taken into account twice.

In the next section we will shortly describe our model calculations;
Sect.~3 contains the comparison with the globular clusters M15, M68
and M92. We have concentrated on these three very old clusters,
because the intention of this {\sl Letter} is to reconcile cluster and
cosmological ages. In a forthcoming paper we will discuss a larger set
of clusters in a broader context. The conclusions will follow in the
last section, as usual.

\section{Models}

All the evolutionary calculations presented in the paper
have been performed with the Frascati
Raphson Newton Evolutionary Code (FRANEC) whose general features and
physical inputs have already been described in previous papers (see
e.g. Chieffi \& Straniero 1989).
We have adopted the OPAL opacity tables
(Rogers \& Iglesias 1992; Rogers \& Iglesias 1995; Iglesias \& Rogers
1995; Rogers 1995,
private communication) combined with
the molecular opacities by Alexander \& Ferguson (1994). 
More precisely, two sets of opacity tables 
(Alexander, Iglesias and Rogers, private communication) 
were used,
one for a scaled solar metal mixture (Grevesse \& Noels 1993)  
and one for the same heavy element fraction $Z$, but with the
$\alpha$-elements being enhanced relative to the iron within the metal 
group. In particular,
oxygen is enhanced by $[O/Fe] = 0.5$, and the other
$\alpha$-elements by similar, but slightly varying amounts,
according to the observed values in low metallicity stars (see e.g.
Wheeler et al. 1989). 
It should be emphasized that the 
metal mixtures for both the low (molecular)- and
high-temperature opacity tables are exactly the same.
Thus we were able to compare consistently
stellar models with the same total metallicity but with different
internal distributions of the heavy elements. 
In the high density region, which is not covered by the OPAL opacities 
and in which the dominant source of opacity is due to the electron 
conduction, we used the opacity coefficients by
Itoh et al.\ (1983).

As for the equation of state (EOS), the updated OPAL EOS
(Rogers 1994; Rogers, Swenson \& Iglesias 1996) has been used, upon
which the OPAL opacities also rest. In regions, where the OPAL EOS is
not available, we supplemented it by the EOS described by Chieffi \& Straniero
(1989; for $T < 5000 $ K) and Straniero (1988; for degenerate He cores
and the central regions of non-degenerate ZAHB He cores).
We have verified that the
transition from the OPAL to the supplementary EOS is smooth and without
discontinuities.

Stellar models
were evolved with a total metallicity of Z=0.0002 and Z=0.0004, a
scaled solar and an $\alpha$-enhanced metal distribution,  
and helium abundances Y=0.23 and Y=0.24. These mixtures were
chosen to match the observed values. For each given chemical
composition, models
with masses ranging from 0.7 to 1 M$_\odot$ were 
evolved from the Zero Age Main Sequence
(ZAMS) up to the Red Giant Branch (RGB) until a luminosity of
Log L/L$_\odot \approx$ 1.7. We have chosen 0.7 M$_\odot$ as a lower mass
limit for our models in order to have, at least for the evolution
up to the turn-off point (TO), all the structure of our models covered by
the OPAL EOS (see also the discussion in CK95).
In all models a mixing length of 1.6
was used; we checked that with this value of the mixing length 
the observational data of Frogel et al. (1983) for the temperatures of the RGB
are reproduced. Isochrones were constructed by interpolating
among evolutionary tracks for each selected chemical
composition.

The 0.8 M$_\odot$ models have been evolved until the core helium flash to
derive the mass of the helium core and the envelope He abundance after 
the first dredge-up; for this mass the stellar age at
the He--flash is of the order of 13 Gyr, roughly
representative of our derived age for metal-poor globular
clusters (see the following section). 
For the ZAHB models, we employed the He core mass and the envelope 
chemical composition at the flash and  
computed an initial set of He-burning models 
with different masses of the H-rich envelopes.  Following
Castellani et al.\ (1991), we assumed as ZAHB structures models
already evolved by 1 Myr, which should represent quite accurately
the theoretical counterparts of the lower luminosity boundaries of the
observed HBs.
  
The color transformations and bolometric corrections used to transform
theoretical temperatures and luminosities into magnitudes in the UBV
system came from Buser \& Kurucz (1978, 1992, hereinafter BK78,
BK92). We adopted the BK92 transformations for $T\leq6000$ K, and the
BK78 at higher temperatures according to the suggestions by BK92. For each 
metallicity, the two sets have two temperatures in common 
(5500 and 6000 K, resp.), 
and we have shifted the $B-V$ values by BK78 in order to match the
corresponding ones by BK92 at these two temperatures (see also De
Santis 1996).
This set of transformations, as discussed by BK92, is highly
homogeneus, and covers all the evolutionary stages displayed in the
Color Magnitude Diagrams of the studied clusters.  

As a comment, we add that we confirm a result obtained
by Salaris, Chieffi \& Straniero (1993): Comparing the $\alpha$-enhanced 
isochrones and ZAHBs with the scaled solar
ones for the same total metallicity, we find that the
$\alpha$-enhanced models are well reproduced by scaled solar ones . In
contrast to Salaris et al.\ (1993) we had 
access to opacity tables reflecting the $\alpha$-enhancement for all 
temperatures. Note that this equivalence holds for low metallicities only 
(Weiss, Peletier \& Matteucci 1994)

\section{Cluster fits}

The $\alpha$-enhanced isochrones
described in the previous section  have  been  used 
for  determining  the  ages of metal   poor   galactic 
globular clusters. We have selected M92 (NGC6341), M68 (NGC4590) and 
M15 (NGC7078) as being representative of the old metal-poor globular 
cluster population.
These three clusters  have  been  extensively 
studied in the past, and according to various  authors  (see  e.g. 
Vandenberg, Bolte \& Stetson  1990;
Straniero  \&  Chieffi  1991; Chaboyer, Sarajedini \& Demarque 1992;  
Salaris et al.\ 1993) they are among the oldest clusters in 
the Galaxy; therefore their age put strong constraints on the age of
the universe and on the Hubble constant. 
The ages derived 
in the past years, adopting different theoretical 
isochrones and different age indicators (as the $\Delta(V)$  
and the $\Delta(B-V)$, see the discussion in Salaris et al.\ 1993), 
range between 15 and 20 Gyr for M15 and M92, and between 13 
and 19 Gyr for M68
(see e.g. Straniero \& Chieffi 1991; Carney
et al. 1992; Chaboyer et al. 1992). 
Very recently, by using the OPAL equation of state
in computing the theoretical evolutionary models, Mazzitelli, D'Antona \& 
Caloi (1995)
found an age of around 14 Gyr for M68, while CK95
obtained 17, 15 and 13 Gyr for M92, M15 and M68, respectively.
All these authors use the $\Delta(V)$ age indicator in order
to derive the age of the clusters. 
In our analysis 
we have adopted the recent photometry  by  Carney et al.\  
(1994) together with the fiducial line from the work by Stetson \& 
Harris (1987) for the cluster M92; in the case of M68 and  M15  we 
have used the data  published 
by Walker (1994)  and by  Hasley  \& 
Christian (1994), respectively. The values of  $[M/H]$  (taking 
into    account    the    observed    overabundance     of     the 
$\alpha$-elements) adopted for the clusters
come  from  the  paper  by  Salaris  \&  Cassisi   (1996),   where 
spectroscopical determinations of $[\alpha/Fe]$ and  $[Fe/H]$  are 
collected for a sample of 22 globular clusters.  

\begin{figure}[t]
\epsscale{0.60}
\plotone{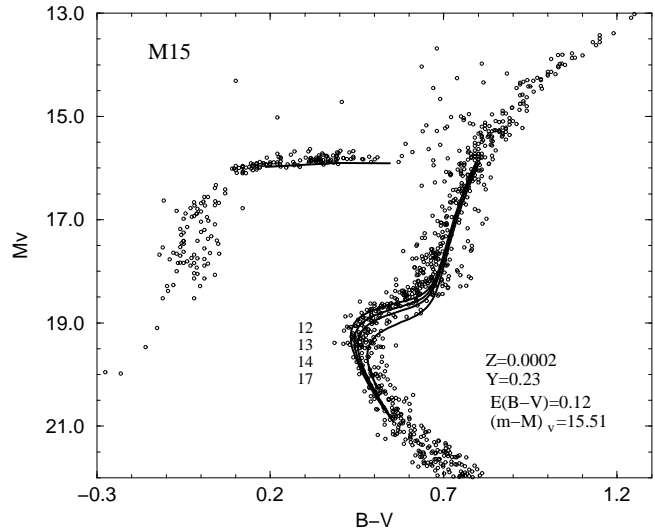}
\caption{Isochrones for different ages (given in Gyrs in the figure)
fitted to the CMD of M15. Composition, reddening and distance modulus
are displayed in the lower right corner. For comparison, an isochrone
of 17 Gyr is shown as well. See text for more details.}
\label{M15.fig}
\end{figure}

\begin{figure}[b]
\epsscale{0.60}
\plotone{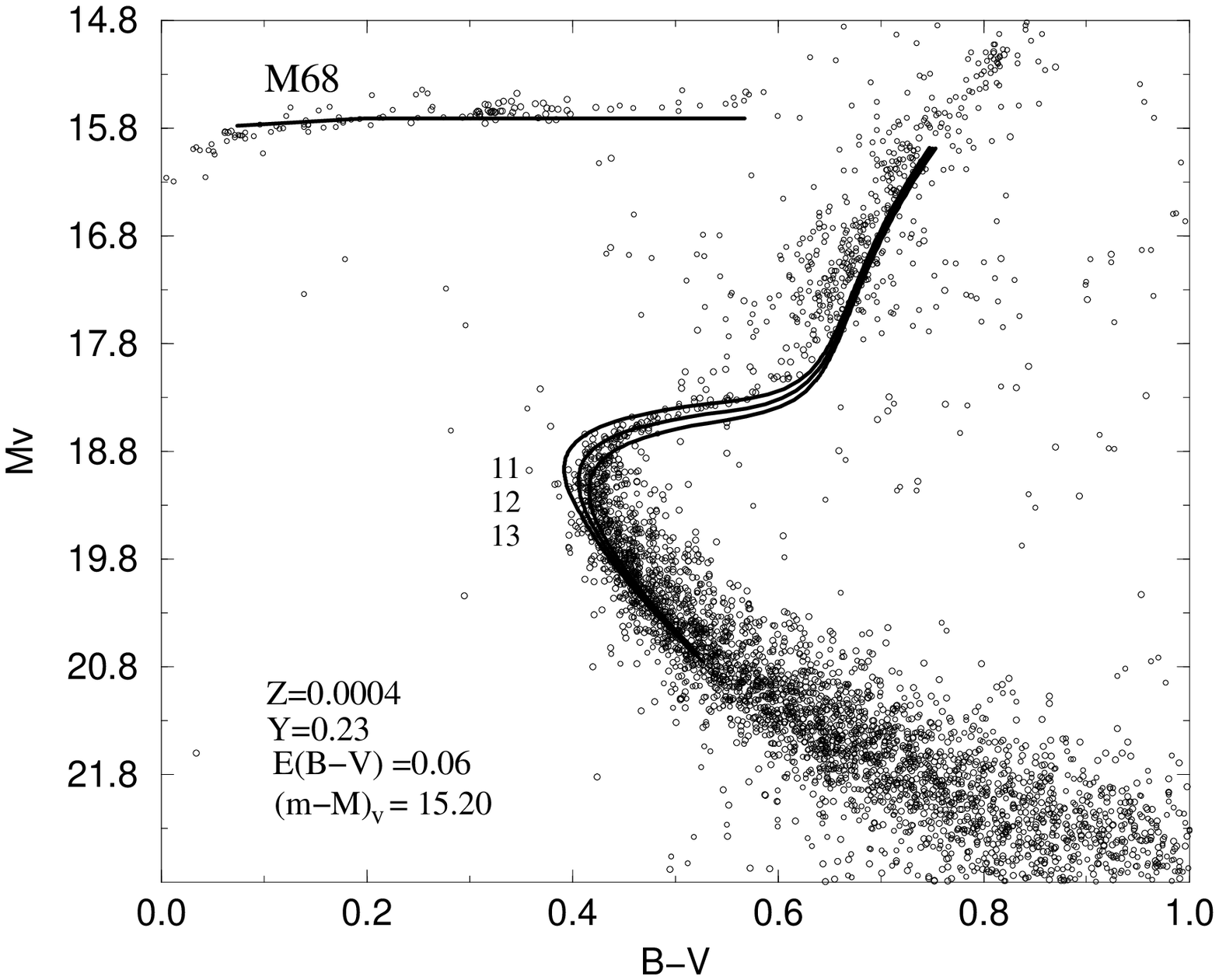}
\caption[]{Same as Fig.~\ref{M15.fig}, but for M68.}
\label{M68.fig}
\end{figure}

\begin{figure}
\epsscale{0.60}
\plotone{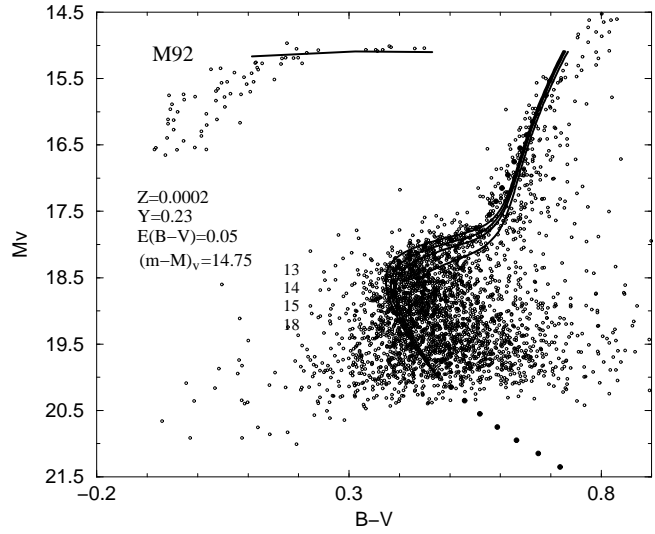}
\caption[]{Same as Fig.~\ref{M15.fig}, but for M92. The filled dots are
the fiducial line of Stetson \& Harris 1988. For comparison, an
isochrone for 18 Gyr (the preferred age of M92 in the past) is shown.}
\label{M92.fig}
\end{figure}

In Figs.~\ref{M15.fig}, \ref{M68.fig} and \ref{M92.fig} the fit of the
theoretical isochrones to the  
observational data of the three clusters  is  displayed. 
On the data by Carney et al. (1992) for M92 the main line
by Stetson \& Harris (1988) for the MS and RGB loci
has been superposed, while in the case of M68,
for a sake of clarity,
we have displayed the photometry from only two fields among the  
long exposure frames, and the data from all the short exposure frames.
We  have 
derived the distance moduli of the clusters and  their  reddening 
by simultaneously superposing our ZAHB location to the lower envelope of 
the observed horizontal part of the clusters HBs in 
the region of the RR-Lyrae stars, and  by  fitting 
with the isochrones the observed MS and RGB loci. 
Once distance modulus and  reddening are fixed, the fit to the TO 
luminosity provides the age of the cluster.
By adopting Y=0.23 and a  metallicity  Z=0.0002  for M92 and M15 
(corresponding approximatively to the estimated values [M/H]=-2.04  and  -2.09)  we 
derived from the fit an age  of 13--14 Gyr
for M92, 12--13 Gyr for M15.
In the case of M68 the estimated value of  [M/H]=-1.78  corresponds 
approximatively to a metallicity
Z=0.0004. The derived age is 12--13 Gyr, too.
Note that our ages are again lower as compared to CK95. 
For the same cluster, we have tested the influence 
of a slightly higher original helium content (Y=0.24) and of a variation 
of the metallicity by a factor of 2 (Z=0.0002). In the first case we 
obtained a reduction  of the age by slightly less then 1 Gyr, 
in the latter case an increase by the same amount; both findings are
in agreement with Salaris et al.\ (1994). 

We have also derived the age of M68 using the 
$\alpha$-enhanced isochrones transformed to the observational
plane by adopting the transformations by Kurucz (1992).
These transformations cover homogeneously all the range of effective
temperatures and gravities from the MS to
the HB evolutionary phase. 
Since there is some
concern about the Kurucz (B-V)-Teff relation, especially for RGB
metal poor stars
(see e.g. Bell et al.\ 1994; McQuitty et al\ 1994; 
Gratton et al. 1996; D'Antona \& Mazzitelli 1996; Alexander et al.\ 1996),
we have only considered the difference in luminosity between the observational
ZAHB in the region of the RR-Lyrae stars 
and the TO of the cluster, obtaining an age reduced by almost 1 Gyr
with respect to the fit displayed in Fig.~\ref{M68.fig}. 

\clearpage

\section{Discussion and conclusions}

In this paper we have reexamined the age of three of the oldest globular 
clusters (M15, M68, M92). Changes to earlier work included the use of the 
latest OPAL EOS and opacity tables, supplemented by low-temperature 
opacity tables for {\em exactly} the same compositions. In particular, 
$\alpha$-enhancement within the metals could be taken into account. For 
appropriate chemical compositions both isochrones {\em and} ZAHB models were 
calculated and cluster ages were derived from the $\Delta(V)$  
difference between TO and ZAHB. We find that the age of the clusters is 
about 13 Gyr, and that the three clusters are practically coeval. These 
ages are lower by up to 4--5 Gyr as compared to earlier results,
e.g. Salaris et al.\ (1993). They made use of the bolometric
corrections of Vandenberg \& Bell (1985) for the TO, while we switched
to BK92 and BK78. This, as we have checked, lowers the ages already by
almost 2 Gyrs. 
The additional age reduction results from the new OPAL EOS. Our 
results confirm and extend those of CK95 and Mazzitelli 
et al.\ (1995). The further reduction as compared to CK95 
is due to our additional ZAHB calculations. 

The derived ages {\em can further be reduced} by the following means: a) 
the use of a higher helium content, justified by Big Bang Nucleosynthesis 
results on the best-fit predicted primordial helium content of 0.247 (Hata et
al.\ 1996) or by observations including a large systematic error
(Olive \& Steigman 1995);
b) the use of the new Kurucz (1992) bolometric corrections; we did not 
use them because of the problems encountered with RGB colors; 
c) the inclusion of diffusion (Chaboyer et al.\ 1992). 
Each of these factors reduces 
the ages by another 0.5 Gyr at least, such that an age of the oldest 
globular clusters of 12 Gyr seems to be in reach.

Interestingly, the mean age obtained for the
three clusters is almost coincident with the age of the Disk (10--12
Gyr) obtained
by Hernanz et al. (1994) by means of the luminosity function of the
white dwarfs in the solar neighbourood, thus implying that the
Galactic Disk began to form without time delay with respect to the halo. 

In a forthcoming paper we will present extended results for a large 
sample of clusters.  
The bottom line of the present {\sl Letter} is that with updated
physics 
the oldest globular clusters are only 13 Gyr (or less) old. This
reduces the ``Age Conflict'' drastically. Stellar evolution 
theorists have done the first step. It is now for 
the cosmologists to confirm that $H_0 \approx 50$.

\acknowledgments
We are grateful to Drs.~Alexander and Rogers for providing us with
their latest opacity tables prior to publication and for computing
special tables for our purposes. It is a pleasure to thank V.
Castellani for helpful discussions.

\end{document}